\documentclass[11pt, a4paper]{article}

\usepackage{amsmath,latexsym}
\usepackage{epsfig}
\usepackage[cp1251]{inputenc}
\usepackage[T2A]{fontenc}
\newcommand{\be}{\begin{equation}}
\newcommand{\ee}{\end{equation}}
\newcommand{\bea}{\begin{eqnarray}}
\newcommand{\eea}{\end{eqnarray}}

\newtheorem{theorem}{Theorem}

\newcommand{\benumerate}{\begin{enumerate}}
\newcommand{\eenumerate}{\end{enumerate}}

\newcommand{\bitemize}{\begin{itemize}}

\newcommand{\eitemize}{\end{itemize}}

\begin{document}

\title{Dispersive deformations of Hamiltonian systems of hydrodynamic type in $2+1$ dimensions}
\author{E.V. Ferapontov, V.S. Novikov and N.M.  Stoilov }
    \date{}
    \maketitle
    \vspace{-7mm}
\begin{center}
 Department of Mathematical Sciences \\ Loughborough University \\
Loughborough, Leicestershire LE11 3TU \\ United Kingdom \\[2ex]
e-mails: \\[1ex] \texttt{E.V.Ferapontov@lboro.ac.uk}
\\[1ex] \texttt{V.Novikov@lboro.ac.uk}\\[1ex]
\texttt{N.M.Stoilov@lboro.ac.uk}
\end{center}

\bigskip

{\it Dedicated to Professor Boris Dubrovin on the occasion of his 60th birthday}

\begin{abstract}

We develop a theory of integrable dispersive deformations of $2+1$ dimensional Hamiltonian  systems of hydrodynamic type following the scheme proposed by Dubrovin and his collaborators in $1+1$ dimensions. Our results show that the multi-dimensional situation is far more rigid, and generic   Hamiltonians are not deformable. As an illustration we discuss a particular class of two-component Hamiltonian systems, establishing the  triviality of first order deformations and classifying Hamiltonians possessing nontrivial deformations of the second order.

\bigskip


\bigskip

\end{abstract}

\newpage

\section{Introduction}

Deformation theory of $1+1$ dimensional Hamiltonian systems  has been thoroughly investigated by Dubrovin and his collaborators in \cite{Dubrovin3, Dubrovin4, Dubrovin5, Dubrovin6, Dubrovin7}: given a Hamiltonian system of hydrodynamic type,
\begin{equation}
u^i_t=\{u^i, H_0\}=P^{ij}\ \delta H_0/\delta u^j,
\label{1+1}
\end{equation}
$i, j=1, \dots, n$, where $P^{ij}=\epsilon^i\delta^{ij}d/dx$ is  the Hamiltonian operator  and $H_0=\int h(u)\ dx$ is a Hamiltonian with the density $h(u)$, one looks for deformations of the form
\begin{equation}
 H= H_0+\epsilon H_1+\epsilon^2 H_2+\dots
\label{Hdef}
\end{equation}
where  the density of $H_i$ is  assumed to be a homogeneous polynomial of degree $i$ in the $x$-derivatives of $u$.  Here the Hamiltonian operator $P^{ij}$ can be assumed undeformed due to the general results of \cite{Getzler, Magri}. Deformation (\ref{Hdef})  is called integrable (to  the order $\epsilon^m$) if any hydrodynamic Hamiltonian $F_0=\int f(u)\ dx$ commuting with $H_0$ can be deformed in such a way that $\{H, F\}=0 \ ({\rm mod} \ \epsilon^{m+1})$. It is  assumed that $H_0$ generates an integrable system of hydrodynamic type \cite{Tsarev0, Tsarev, Dubrovin2}: any system of this kind possesses an infinity of commuting Hamiltonians $F_0$ parametrised by  $n$ arbitrary functions of one variable. The classification of integrable deformations   is performed modulo canonical transformations  of the form 
\begin{equation}
H\to H+\epsilon \{K, H\}+\frac{\epsilon^2}{2}\{K, \{K, H\}\}+\dots
\label{can}
\end{equation}
where $K$ is any functional of the form (\ref{Hdef}).  The richness  of this deformation scheme is due to the following basic facts:
\begin{itemize}
\item The variety of integrable `seed' Hamiltonians $H_0$ is parametrised by $n(n-1)/2$ arbitrary functions of two variables;

\item For a fixed integrable  Hamiltonian $H_0$,  the deformation procedure  introduces extra arbitrary functions of one variable known, in bi-Hamiltonian context, as `central invariants'.  One should point out that it is still an open problem to extend a deformation, for arbitrary values of these functions, to all orders in the deformation parameter $\epsilon$. 
\end{itemize}

The main goal of this paper is to discuss the analogous deformation scheme in $2+1$ dimensions. One again starts with the Hamiltonian system (\ref{1+1}) where $P^{ij}$ is a $2$-dimensional Hamiltonian operator of hydrodynamic type,
see \cite{Dubrovin1, Mokhov1, Mokhov2} for the general theory and classification results. 
In the two-component case there exist only three types of  such operators:  the first two of them can be reduced to  constant-coefficient forms,  
\begin{equation*}
P= \left(
\begin{array}{cc}
d/dx&0\\
0 & d/dy
\end{array}
\right),
~~~ P= \left(
\begin{array}{cc}
0&d/dx\\
d/dx & d/dy
\end{array}
\right),
\end{equation*}
while the third one is essentially non-constant,
\begin{equation*}
P= \left(
\begin{array}{cc}2v&w\\
w & 0
\end{array}
\right)\frac{d}{dx}+
 \left(
\begin{array}{cc}
0&v\\
v & 2w
\end{array}
\right)\frac{d}{dy}+
 \left(
\begin{array}{cc}
v_x&v_y\\
w_x & w_y
\end{array}
\right),
\end{equation*}
here $v, w$ are the dependent variables. We will refer to them as Hamiltonian operators of type I, II and III, respectively. To be specific, we will concentrate on   case II. The corresponding Hamiltonian systems take the form
\begin{equation}
\left(
  \begin{array}{c}
  v \\
    w \\
  \end{array}
\right)_t    =\left(
  \begin{array}{cc}
    0 & d/dx \\
    d/dx & d/dy \\
  \end{array}
\right)\left(
                                            \begin{array}{c}
                                              \delta H_0/\delta v \\
                                           \delta H_0/\delta w     \\
                                            \end{array}
                                          \right),
\label{H1}
\end{equation}
$H_0=\int h(v, w)\ dxdy$, or, explicitly,
$$
v_t=(h_w)_x,  ~~~
 w_t=(h_v)_x+(h_w)_y.
$$
We will be looking at deformations of the form (\ref{Hdef}) where the density of  $H_i$ is a homogeneous polynomial of degree $i$ in the $x$- and $y$-derivatives of $v$ and $w$. 
The Hamiltonian operator will be assumed undeformed (although we are not aware of  any results establishing the triviality of  Poisson cohomology in higher dimensions). Since a system of the form (\ref{H1}) does not possess any nontrivial conservation laws of hydrodynamic type other than the Casimirs and the Hamiltonian, the definition of integrability needs to be  modified. Thus, the `seed'  system (\ref{H1}) will be called integrable if it possesses infinitely many hydrodynamic reductions \cite{ GibTsa96,  Fer1}. This requirement imposes strong constraints on the Hamiltonian density $h(v, w)$, providing an efficient classification criterion  (see Sect. 2 for more details). Following \cite{FM, FMN}, a deformation of $H_0$ will be called integrable (to the order $\epsilon^m$) if it inherits all hydrodynamic reductions of the seed system (\ref{H1}) to the same order (the deformation procedure is outlined in Sect. 3). The main features of the $2+1$ dimensional deformation scheme can be summarised as follows:

\begin{itemize}
\item The variety of integrable `seed' Hamiltonians $H_0$ is finite dimensional. 

\item {\it  Generic} integrable Hamiltonians $H_0$ possess no nontrivial deformations. 
\end{itemize}
 Nevertheless, there exist deformable (non-generic) Hamiltonians. 
 
 \medskip
 
 \noindent {\bf Example 1.} Let $H_0=\int \frac{w^2}{2}+f(v)\ dxdy$. In this case the integrability  conditions reduce to a single fourth order ODE,  $f''''f''=f'''^2$, so that without any loss of generality one can set $f(v)=e^v$.  Modulo canonical transformations, this Hamiltonian  possesses a unique integrable dispersive  deformation of the form 
  $$
 H=\int  \frac{w^2}{2}+f(v)-\frac{\epsilon^2}{3!}f'''v_{x}^2+O(\epsilon^4)\ dxdy.
 $$
For $f(v)=e^v$  it can be rewritten in the equivalent form
 $$
 H=\int  \frac{w^2}{2}+e^v+\frac{\epsilon^2}{3!}e^v v_{xx}+O(\epsilon^4)\ dxdy.
 $$
 It is quite remarkable that this deformation can be extended to all orders in the deformation parameter $\epsilon$, providing a Hamiltonian formulation of the $2D$ Toda system,
$$
H=\int   \frac{w^2}{2}+ exp\left(v+\frac{\epsilon^2}{3!}v_{xx}+\frac{\epsilon^4}{5!}v_{xxxx}+\dots \right)\ dxdy,
$$
see Sect. 6 for further details. 

\medskip

 \noindent {\bf Example 2.} Let $H_0=\int \alpha \frac  {v^2}{2}+\beta vw+ f(w)\ dxdy$. Here the integrability conditions reduce to a single fourth order ODE,  $f''''f''(\alpha f''-\beta^2)=f'''^2(3\alpha f''-2\beta^2)$.   Modulo canonical transformations, this Hamiltonian possesses a unique integrable dispersive deformation of the form
   $$
 H=\int  \alpha \frac{v^2}{2}+\beta vw+f(w)+{\epsilon^2}f'''\left(-\frac{\alpha^2}{f''} v_x^2+\frac{\beta^2}{f''} v_y^2+2\alpha w_x^2+2\beta v_yw_y+f''w_y^2\right)+O(\epsilon^4)\ dxdy.
 $$
Although for $\beta=0$ the Hamiltonian $H_0$ gives rise to the  dispersionless KP (dKP) equation,  the deformation presented here is not equivalent to the full KP equation:  see Sect. 7 for further discussion.
 
It will be demonstrated  (Theorem 2 of Sect. 5) that, modulo certain equivalence transformations, these  two examples exhaust the list of   Hamiltonians of type II which possess nontrivial integrable deformations to the order $\epsilon^2$. In Sect. 4 we prove the triviality of $\epsilon$-deformations. The structure of $\epsilon^2$-deformations is analysed in Sect. 5.  Deformations of Hamiltonians of type II/III are discussed in Sect. 8/9. 

\section{Classification of integrable Hamiltonian densities of type II}

In this section we review the classification of integrable   Hamiltonian systems of the form (\ref{H1}). 
Following \cite{Fer1, FOS} we require the existence of $N$-phase solutions of the form 
\begin{equation}
v = v(R^1, R^2, \dots, R^N), ~~~ w = w(R^1, R^2, \dots, R^N),
\label{vw}
\end{equation}
 where the phases $R^i(x, y, t)$ satisfy the commuting equations
\begin{equation}
R^i_t = \lambda^i (R) R_x^i, ~~~ R^i_y = \mu^i (R) R_x^i;
\label{R}
\end{equation}
recall that the assumption of commutativity imposes the following restrictions on  the chraracteristic speeds  $\lambda^i$ and $\mu^i$:
\begin{equation}
\frac{\partial_j\lambda^i}{\lambda^j-\lambda^i}=\frac{\partial_j\mu^i}{\mu^j-\mu^i},
\label{comm}
\end{equation}
$\partial_j=\partial/\partial_{R^j}, \ i\ne j$, see \cite{Tsarev}. Equations (\ref{R}) are said to define an $N$-component  hydrodynamic reduction of the original system (\ref{H1}). It was observed in \cite{Fer1} that the requirement of the existence of such reductions imposes strong constraints on the original system (\ref{H1}), and provides an efficient classification criterion.  Recall that the key property is the existence of three-component reductions: in this case one also has $N$-component reductions for arbitrary $N$. This property is reminiscent of the three-soliton condition in the theory of integrable systems. On the contrary, the existence of one- or two-component reductions is a common phenomenon which is not generally related to the integrability (at least for two-component systems as in the present paper).  As shown in \cite{FOS},  the requirement of  existence of three-component  reductions leads to a system of fourth order PDEs for the  Hamiltonian density $h(v, w)$, which constitute the integrability conditions:
\begin{equation}
\begin{array}{c}
 h_{ww}(h_{vv}h_{ww}-h_{vw}^2)h_{vvvv} = 
-6h_{vw}^2 h_{vvw}^2  +3 h_{ww} h_{vv} h_{vvw}^2  
\\
+4 h_{vw}^2h_{vww} h_{vvv}  -  2h_{ww} h_{vw} h_{vvw} h_{vvv}+
h_{ww}^2h_{vvv}^2,\\
\  \\
h_{ww}(h_{vv}h_{ww}-h_{vw}^2)h_{vvvw} =  
-3h_{vw}^2 h_{vww}h_{vvw}  + 3h_{ww} h_{vww} h_{vvw} h_{vv} 
\\
- 3h_{ww} h_{vw} h_{vvw} ^2
+ h_{vw}^2 h_{vvv} h_{www} + h_{ww} h_{vw} h_{vww} h_{vvv} + h_{ww}^2 h_{vvw} h_{vvv},\\ 
\ \\
h_{ww}(h_{vv}h_{ww}-h_{vw}^2)h_{vvww} = 
-2h_{vw}^2 h_{vww}^2 + 2h_{vv} h_{vww}^2 h_{ww} 
\\
- 3h_{ww} h_{vw} h_{vww} h_{vvw} + h_{ww} h_{vv} h_{www} h_{wvv} 
+ h_{ww} h_{vw} h_{www} h_{vvv} + h_{ww}^2 h_{vww} h_{vvv},\\  
\ \\
h_{ww}(h_{vv}h_{ww}-h_{vw}^2)h_{vwww} = 
-2h_{vw}^2 h_{www} h_{vww} -3h_{ww} h_{vww}^2 h_{vw} 
\\
+ 3h_{ww} h_{vv} h_{www} h_{vww} + h_{ww} h_{vw} h_{www} h_{vvw}
+ h_{ww}^2  h_{www} h_{vvv}, \\  
\ \\

h_{ww}(h_{vv}h_{ww}-h_{vw}^2)h_{wwww} = 
-2h_{vw}^2 h_{www}^2  - 2h_{ww} h_{www} h_{vw}h_{vww} 
\\
- 3h_{ww}^2 h_{vww}^2  + 3 h_{ww} h_{vv} h_{www}^2 
+ 4h_{ww}^2 h_{www} h_{vvw}. 
\end{array} 
\label{4}
\end{equation}
This system is in involution, and is invariant under the $9$-parameter group of  Lie-point symmetries,
\bea
&&{\nonumber}v \rightarrow av +  b,  \\
&&{\nonumber}w \rightarrow pv+ cw + d,  \\
&&{\nonumber}h \rightarrow \alpha h + \beta v + \gamma w +\delta.
\eea
These transformations form the equivalence group of the problem. They preserve the Hamiltonian structure,  and will be used to simplify the classification results. Under the Legendre transformation,
$$
V=h_v, \ W=h_w, \ H=vh_v+wh_w-h, \ H_V=v, \ H_W=w,
$$
the integrability conditions (\ref{4}) simplify to 
\bea
{\nonumber}&&H_{VVVV} = \frac{2 H^2_{VVV}}{H_{VV}}, \\
{\nonumber}&&H_{VVVW} = \frac{2 H_{VVW}H_{VVV} }{H_{VV}}, \\
{\nonumber}&&H_{VVWW} = \frac{2 H^2_{VVW}}{H_{VV}}, \\
{\nonumber}&&H_{VWWW} = \frac{3 H_{VWW}H_{VVW} - H_{WWW}H_{VVV}}{H_{VV}}, \\
{\nonumber}&&H_{WWWW} = \frac{6H^2_{VWW} - 4H_{WWW}H_{VVW}}{H_{VV}}.
\eea
These equations were explicitly solved in \cite{FOS}, leading to the following classification result:

\medskip

\noindent

\begin{theorem} Modulo the natural equivalence group, the  generic  integrable potential  $H(V,W)$ of type II is given by the formula
$$
H = V\ln{\frac{V}{\sigma(W)}},
$$
where $\sigma$ is the Weierstrass sigma-function: $\sigma' / \sigma= \zeta,  \ \zeta' = -\wp, \ \wp^{\prime 2} = 4\wp^3 - g_3$. Its degenerations correspond to
$$
H = V\ln \frac{V}{W}, 
~~~
H = V\ln{V},
~~~
H = \frac{V^2}{2W} + \alpha W^7, 
$$
as well as the following polynomial  potentials:
$$
H = \frac{V^2}{2} + \frac{VW^2}{2} + \frac{W^4}{4}, ~~~
H = \frac{V^2}{2} + \frac{W^3}{6}. 
$$
\end{theorem}
Taking the inverse Legendre transform, one can obtain a complete list of integrable Hamiltonian densities $h(v, w)$. Just to mention a few of them, one gets
$$
h(v, w)=\frac{w^2}{2}+e^v, ~~~ h(v, w)=\frac{v^2}{2}+w^{3/2}, ~~~ h(v, w)=\frac{1}{2}(w+v^2/2)^2, ~~~ h(v, w)=\frac{1}{2}(w+e^v)^2,
$$
etc. 
However, we would prefer to avoid  case-by-case considerations, and work with the full set of integrability conditions (\ref{4}).

\section{Dispersive deformations in $2+1$ dimensions}

Given a Hamiltonian system of the form (\ref{H1}), its deformation $ H= H_0+\epsilon H_1+\dots + \epsilon^m H_m+O(\epsilon^{m+1})$ will be called integrable  (to the order $\epsilon^m$) if both equations (\ref{vw}) and (\ref{R}) defining $N$-phase solutions can be deformed to the same order in  $\epsilon$, in other words, the deformed dispersive system is required to `inherit' all hydrodynamic reductions of its dispersionless limit \cite{FM, FMN}. More precisely, we require the existence of expansions
\begin{equation}
\begin{array}{c}
v = v(R^1, R^2, \dots, R^N)+\epsilon v_1+\dots +\epsilon^m v_m+O(\epsilon^{m+1}), \\
\ \\
 w = w(R^1, R^2, \dots, R^N)+\epsilon w_1+\dots +\epsilon^m w_m+O(\epsilon^{m+1}),
 \end{array}
\label{vw0}
\end{equation}
 where $v_i$ and $w_i$ are assumed to be homogeneous polynomials of degree $i$ in the $x$-derivatives of $R$'s (thus, both $R^i_{xx}$ and $R^i_x R^k_x$  have degree two, etc). Similarly, hydrodynamic reductions (\ref{R}) are deformed as
\begin{equation}
\begin{array}{c}
R^i_t = \lambda^i (R) R_x^i+\epsilon a_1+\dots +\epsilon^m a_m+O(\epsilon^{m+1}), \\
\ \\
R^i_y = \mu^i (R) R_x^i+\epsilon b_1+\dots +\epsilon^m b_m+O(\epsilon^{m+1}),
\end{array}
\label{R0}
\end{equation}
 where $a_i$ and $b_i$ are assumed to be homogeneous polynomials of degree $i+1$ in the $x$-derivatives of $R$'s. We require that the substitution of (\ref{vw0}), (\ref{R0}) into the deformed system (\ref{H1})  satisfies the equations up to the order $O(\epsilon^{m+1})$. This requirement proves to be very restrictive indeed, and imposes strong constraints on the structure of the deformed Hamiltonian $H$. 
 
 \noindent{\bf Remark.} Expansions (\ref{vw0})-(\ref{R0}) are invariant under Miura-type transformations of the form
 $$
 R^i\to R^i+\epsilon r_1+\epsilon^2 r_2+\dots,
 $$
 where $r_i$ denote terms which are polynomial of degree $i$ in the $x$-derivatives of $R$'s. These transformations can be used to simplify calculations. For instance, working with one-phase solutions one can assume that $v$ remains undeformed. Similarly, working with two-phase solutions one can assume that both $v$ and $w$ remain undeformed. For three-phase solutions this normalisation still leaves some extra Miura-freedom which can be used to simplify expressions for  $a_i$ and $b_i$ (to the best of our knowledge there exist no general theory of normal forms under Miura-type transformations).

\section{Triviality of $\epsilon$-deformations}

In this section we prove that  all $\epsilon$-deformations are trivial and can be eliminated by an appropriate canonical transformation. Thus, we consider deformations of the form
\begin{equation}
\left(
  \begin{array}{c}
  v \\
    w \\
  \end{array}
\right)_t    =\left(
  \begin{array}{cc}
    0 & d/dx \\
    d/dx & d/dy \\
  \end{array}
\right)\left(
                                            \begin{array}{c}
                                              \delta H/\delta v \\
                                             \delta H/ \delta w \\
                                            \end{array}
                                          \right)
\label{H11}
\end{equation}
where
$$
H=\int h(v, w)+\epsilon (av_x+bv_y+pw_x+qw_y)+O(\epsilon^2) \ dxdy.
$$
Here $a, b, p, q$ are functions of $v$ and $w$. 
We require that all $N$-phase solutions (\ref{vw}) can be extended to the order $\epsilon$, 
\begin{equation}
v = v(R^1, R^2, \dots, R^N)+\epsilon v_1+O(\epsilon^2), ~~~ w = w(R^1, R^2, \dots, R^N)+\epsilon w_1+O(\epsilon^2),
\label{vw1}
\end{equation}
where $v_1$ and $w_1$ are polynomials of order one in the $x$-derivatives of $R$'s. Similarly, hydrodynamic reductions (\ref{R}) are deformed as
\begin{equation}
R^i_t = \lambda^i (R) R_x^i+\epsilon a_1+O(\epsilon^2), ~~~ R^i_y = \mu^i (R) R_x^i+\epsilon b_1+O(\epsilon^2),
\label{R1}
\end{equation}
where $a_1$ and $b_1$ are polynomials of order two in the $x$-derivatives of $R$'s. We thus require that  relations (\ref{vw1}), (\ref{R1}) satisfy the original system (\ref{H11})  up to the order $O(\epsilon^2)$.

It was verified by a direct calculation that all one- and two-component reductions can be deformed in this way, for any $a, b, p, q$ and any density $h(v, w)$, not necessarily integrable. On the contrary, the requirement of the inheritance of three-component reductions (recall that the existence of three-component reductions forces $h(v, w)$ to satisfy the integrability conditions (\ref{4})), is nontrivial, and leads to the  following single relation:
\begin{equation}
\left(\frac{h_{vv}N_w-h_{vw}(M_w-N_v)}{h_{vv}h_{ww}-h_{vw}^2}\right)_w=\left(\frac{h_{vw}N_w-h_{ww}(M_w-N_v)}{h_{vv}h_{ww}-h_{vw}^2}\right)_v,
\label{Rel1}
\end{equation}
here  $M=(a_w-p_v)/h_{ww}, \ N=(b_w-q_v)/h_{ww}$. It remains to show that the relation (\ref{Rel1}) is necessary and sufficient for the existence of a canonical transformation of the form
$$
H\to H+\epsilon\{K, H\}+O(\epsilon^2),
$$
with $K=\int k(v, w)\  dxdy$, which eliminates all $\epsilon$-terms. Since the density of the functional $H+\epsilon\{K, H\}$ is given by the formula
$$
h(v, w)+\epsilon  (av_x+bv_y+pw_x+qw_y)+\epsilon (k_v, k_w)\left(
  \begin{array}{cc}
    0 & d/dx \\
    d/dx & d/dy \\
  \end{array}
\right)
\left(
  \begin{array}{c}
  h_v \\
    h_w \\
  \end{array}
\right)+O(\epsilon^2)=
$$
$$
h(v, w)+\epsilon  (Av_x+Bv_y+Pw_x+Qw_y)+O(\epsilon^2),
$$
where
$$
A=a+k_vh_{vw}+k_wh_{vv}, ~~ B=b+k_wh_{vw}, ~~ P=p+k_vh_{ww}+k_wh_{vw}, ~~ Q= q+k_wh_{ww},
$$
the conditions that $\epsilon$-terms are trivial (form a total derivative),  take the form $A_w=P_v, \ B_w=Q_v$. This
leads to the following linear system for $k(v, w)$:
$$
k_{vv}h_{ww}-k_{ww}h_{vv}=a_w-p_v, ~~~ k_{vw}h_{ww}-k_{ww}h_{vw}=b_w-q_v.
$$
The compatibility conditions of these equations for $k$  can be obtained by introducing the auxiliary variable $p$ via the relation $k_{ww}=ph_{ww}$, and solving for the remaining second order derivatives of $k$,
$$
k_{vv}=M+ph_{vv}, ~~~ k_{vw}=N+ph_{vw}, ~~~ k_{ww}=ph_{ww}.
$$
 Cross-differentiating and solving for $p_v$ and $p_w$ we obtain
$$
p_v=\frac{h_{vv}N_w-h_{vw}(M_w-N_v)}{h_{vv}h_{ww}-h_{vw}^2}, ~~~ p_w=\frac{h_{vw}N_w-h_{ww}(M_w-N_v)}{h_{vv}h_{ww}-h_{vw}^2}.
$$
Ultimately, the compatibility condition $p_{vw}=p_{wv}$ gives the required relation (\ref{Rel1}), thus finishing the proof.

\section{Reconstruction of $\epsilon^2$-deformations}

In this section we analyse the structure of  $\epsilon^2$-deformations. The result of the previous section allows us to set  all $\epsilon$-terms equal to zero. Thus, we consider deformations of the form
\begin{equation}
\left(
  \begin{array}{c}
  v \\
    w \\
  \end{array}
\right)_t    =\left(
  \begin{array}{cc}
    0 & d/dx \\
    d/dx & d/dy \\
  \end{array}
\right)\left(
                                            \begin{array}{c}
                                              \delta H/\delta v \\
                                             \delta H/ \delta w \\
                                            \end{array}
                                          \right)
\label{H12}
\end{equation}
where
$$
H=\int h(v, w)+\epsilon^2 h_2(v, w, v_x, w_x, v_y, w_y)+O(\epsilon^3) \ dxdy.
$$
Here $h_2$ is assumed to be of  second order in the $x$- and $y$-derivatives of $v$ and $w$, 
$$
h_2=f_1v_x^2+f_2v_y^2+f_3w_x^2+f_4w_y^2+f_5v_xw_x+f_6(v_xw_y+v_yw_x)+f_7v_yw_y+f_8v_xv_y+f_9w_xw_y,
$$
where  $f_1, \dots, f_9$ are   functions of $v$ and $w$. Note that all terms which are linear in the second order derivatives of $v$ and $w$ can be removed via integration by parts. 
Furthermore,  any expression of the form $f(v, w)(v_xw_y-v_yw_x)$ can be omitted, since its variational derivative is identically zero.
We require that all $N$-phase solutions (\ref{vw}) can be extended to the order $\epsilon^2$, 
\begin{equation}
v = v(R^1, R^2, \dots, R^N)+\epsilon^2 v_2+O(\epsilon^3), ~~~ w = w(R^1, R^2, \dots, R^N)+\epsilon^2 w_2+O(\epsilon^3),
\label{vw2}
\end{equation}
where $v_2$ and $w_2$  are polynomials of order two in the $x$-derivatives of $R$'s. Similarly, hydrodynamic reductions (\ref{R}) are deformed as
\begin{equation}
R^i_t = \lambda^i (R) R_x^i+\epsilon^2 a_2+O(\epsilon^3), ~~~ R^i_y = \mu^i (R) R_x^i+\epsilon^2 b_2+O(\epsilon^3),
\label{R2}
\end{equation}
where $a_2$ and $b_2$  are polynomials of order three in the $x$-derivatives of $R$'s. We thus require that relations (\ref{vw2}), (\ref{R2}) satisfy the deformed system (\ref{H12})  up to the order $O(\epsilon^3)$.
The classification is performed modulo canonical transformations of the form
$$
H\to H+\epsilon\{K, H\}+O(\epsilon^3),
$$
here $K=\epsilon \int  (av_x+bv_y+pw_x+qw_y)\  dxdy$. Note that the density of the functional $\epsilon\{K, H\}$ is given by the following formula (set $m=a_w-p_v, \ n=b_w-q_v$):
$$
\epsilon^2   (\delta K/\delta v_, \ \delta K/\delta w)\left(
  \begin{array}{cc}
    0 & d/dx \\
    d/dx & d/dy \\
  \end{array}
\right)
\left(
  \begin{array}{c}
  h_v \\
    h_w \\
  \end{array}
\right)+O(\epsilon^3)=
$$
$$
\epsilon^2m(h_{vv}v_x^2+h_{vw}v_xv_y+h_{ww}v_xw_y-h_{ww}w_x^2)+
\epsilon^2n(h_{vv}v_xv_y+h_{vw}v_y^2+h_{ww}v_yw_y-h_{ww}w_xw_y)+ O(\epsilon^3).
$$
Our calculations demonstrate that generic integrable Hamiltonians $H_0$ do not possess nontrivial dispersive deformations. To be precise, these deformations are parametrised by
two arbitrary functions, analogous to  $m$ and $n$ above, which can be eliminated by a canonical transformation. There are cases, however, where dispersive  deformations are parametrised by
two arbitrary functions and a constant.  It is exactly this extra constant  which gives rise to a non-trivial deformation. We  emphasize that canonical transformations can be used from the very beginning to bring the deformation to a `normal form': since $h_{ww}\ne 0$ one can set, say,  $f_6=f_9=0$. This normalisation simplifies all subsequent calculations. Our results can be summarised as follows.

\begin{theorem} A Hamiltonian  $H_0=\int h(v, w)\ dxdy $  of type II  possesses a nontrivial integrable deformation to the order $\epsilon^2$ if and only if, along with the integrability conditions (\ref{4}),  it satisfies the additional differential constraints
$$
h_{vvv}h_{vww}-h_{vvw}^2=0, ~~ h_{vvv}h_{www}-h_{vvw}h_{vww}=0, ~~ h_{www}h_{vvw}-h_{vww}^2=0,
$$
that is, 
\begin{equation}
{\rm rank} \ \left(
\begin{array}{ccc}
h_{vvv} & h_{vvw} & h_{vww} \\
h_{vvw} & h_{vww} & h_{www} 
\end{array}
\right)=1.
\label{rank}
\end{equation}
Modulo equivalence transformations, this gives two types of deformable densities:
$$
h(v, w)=\frac{w^2}{2}+e^v, ~~~~  h(v, w)= \alpha \frac {v^2}{2}+\beta vw+ f(w),
$$
where $f(w)$ satisfies the  integrability condition $f''''f''(\alpha f''-\beta^2)=f'''^2(3\alpha f''-2\beta^2)$.

\end{theorem}

\medskip

\centerline{\bf Proof:}

In contrast to the case of $\epsilon$-corrections where all constraints were coming from deformations of three-component reductions, at the order $\epsilon^2$  the main constraints appear  at the level of one-component reductions already.  Furthermore, it was verified by a direct calculation that  multi-component reductions impose no extra conditions. Since the third order derivative  $h_{www}$ appears as a factor in all deformation formulae, there are two cases to consider.

\noindent {\bf Case 1:} $h_{www}=0$. Then the integrability conditions imply 
 $h_{vww}=0$. The further analysis shows that one has to impose an extra condition, namely $h_{vvw}=0$, otherwise all deformations are trivial. Notice that conditions  $h_{www}=h_{vww}=h_{vvw}=0$ clearly imply (\ref{rank}). Modulo equivalence transformations, this is the case of the Hamiltonian density $h(v, w)=\frac{w^2}{2}+e^v$. Its dispersive deformation is given in Example 1 of the Introduction.

\noindent {\bf Case 1:} $h_{www}\ne 0$. In this case one gets a system of  equations for the coefficients $f_1, \dots, f_9$ which contains $f_4$ as a factor. If $f_4$ equals zero, all deformations are trivial. In the case $f_4\ne 0$ one can express  $f_1, f_2, f_3, f_5, f_7, f_8$ in terms of $f_4, f_6, f_9$. What is left will be a system of two compatible first order PDEs for $f_4$, and a system of additional differential constraints for $h(v, w)$ which coincides with (\ref{rank}).  Solving equations for $f_4$ we obtain a constant of integration which is responsible for non-trivial dispersive deformations.  To find integrable Hamiltonian densities satisfying (\ref{rank}) we set $h_{www}=q, \ h_{vww}=pq$. Then the remaining third order derivatives of $h$ can be parametrised as 
$$
h_{www}=q, ~~ h_{vww}=pq, ~~ h_{vvw}=p^2q, ~~ h_{vvv}=p^3q.
$$
Calculating the compatibility conditions we obtain $p_v=pp_w, \ q_v=(pq)_w$. With this ansatz the integrability conditions (\ref{4}) imply $p$=const so that $q=F(w+pv+c)$ where $f$ is a function of one variable. Thus, $h$ can be represented in the form $h(v, w)=f(w+pv+c)+Q(v, w)$, where $Q(v, w)$ is an arbitrary quadratic form. Modulo the equivalence group any such density can be written in the form
$h(v, w)= \alpha \frac {v^2}{2}+\beta vw+ f(w)$, and the substitution into (\ref{4}) gives a fourth order ODE for $f$. The dispersive deformation of this Hamiltonian  is presented in Example 2.
We believe that both Hamiltonians from Theorem 2 can be deformed to all orders in  $\epsilon$.

\section{Example 1:  deformation of the Boyer-Finley equation}

In this section we discuss the key example where dispersive deformations can be reconstructed explicitly at all orders of the deformation parameter $\epsilon$.  Let us consider  system
(\ref{H1}) with the Hamiltonian density 
$h = \frac{w^2}{2} + e^v,$ 
$$
v_t  = w_x , ~~~
 w_t = e^vv_x+ w_y.
$$
On the elimination of $w$,  it reduces to the Boyer-Finley equation \cite{BF},
\be
\nonumber v_{tt} - v_{ty} = (e^{v})_{xx},
\ee
(the left hand side can be put into the standard form $v_{ty}$ by a linear transformation of $t$ and $y$). An integrable dispersive deformation of this example is closely related to the 2D Toda equation, see \cite{Bla} for an equivalent construction based on the central extension procedure. Let us introduce the auxiliary Hamiltonian system
\begin{equation}
\left(
  \begin{array}{c}
  u \\
    w \\
  \end{array}
\right)_t    =\left(
  \begin{array}{cc}
    0 & \frac{1}{\epsilon} \sinh (\epsilon d/dx) \\
\frac{1}{\epsilon} \sinh (\epsilon d/dx) & d/dy \\
  \end{array}
\right)\left(
                                            \begin{array}{c}
                                              h_u \\
                                              h_w \\
                                            \end{array}
                                          \right),
\label{H2}
\end{equation}
where the Hamiltonian density $h$ is the same as above, $h(u, w)=\frac{w^2}{2} + e^u$ (the exact relation between $u$ and $v$ is specified below). Explicitly, this gives
$$
u_t= \frac{1}{\epsilon} \sinh (\epsilon d/dx) w,  ~~~
 w_t= \frac{1}{\epsilon} \sinh (\epsilon d/dx) e^u+w_y,
$$
which, on elimination of $w$, leads to the integrable 2D Toda equation,
$$
u_{tt}-u_{ty}=\frac{1}{\epsilon^2}( \sinh (\epsilon d/dx) )^2e^u=       \frac{1}{4\epsilon^2}\left(e^{u(x+2\epsilon)}+e^{u(x-2\epsilon)}-2e^{u(x)}\right).
$$
Introducing the change of variables $u\leftrightarrow v$ by the formula
$$
u=\frac{1}{\epsilon}(d/dx)^{-1} \sinh (\epsilon d/dx)v=(d/dx)^{-1} \left(\frac{v(x+\epsilon)-v(x-\epsilon)}{2\epsilon}\right)=v+\frac{\epsilon^2}{3!}v_{xx}+\frac{\epsilon^4}{5!}v_{xxxx}+\dots,
$$
one can verify that the Hamiltonian operator in (\ref{H2}) transforms into the Hamiltonian operator in (\ref{H1}), while the Hamiltonian density $h(u, w)=\frac{w^2}{2} + e^u$ takes the form 
$$
h(v, w)= \frac{w^2}{2} + exp\left(\frac{1}{\epsilon}(d/dx)^{-1} \sinh (\epsilon d/dx)v\right)= \frac{w^2}{2} + exp\left(v+\frac{\epsilon^2}{3!}v_{xx}+\frac{\epsilon^4}{5!}v_{xxxx}+\dots \right)=
$$
$$
 \frac{w^2}{2} +e^v\left(1+\frac{\epsilon^2}{3!}v_{xx}+\frac{\epsilon^4}{5!}(v_{xxxx}+\frac{5}{3}v_{xx}^2)+\dots\right).
$$
This provides the required integrable deformation for the Hamiltonian density $h = \frac{w^2}{2} + e^v$. The $1+1$ dimensional $y$-independent limit of this construction was  discussed in \cite{Dubrovin8}.

\section{Example 2: deformation of the dKP equation}

For $\beta =0, \  \alpha=1$ the  Hamiltonian density from Example 2 takes the form $h(v, w)=  \frac  {v^2}{2}+ f(w)$ where  $f''''f''=3f'''^2$.   The corresponding deformation assumes the form
    $$
 H=\int   \frac{v^2}{2}+f(w)+{\epsilon^2}f'''\left(-\frac{1}{f''} v_x^2+2 w_x^2+f''w_y^2\right)+O(\epsilon^4)\ dxdy.
 $$
 Without any loss of generality one can set $f(w)=\frac{2\sqrt2}{3}w^{3/2}$. In this case the dispersionless system takes the form
  $$
  v_t=(\sqrt{2w})_x, ~~~ w_t=v_x+(\sqrt {2w})_y.
  $$
Introducing the new variable $u=\sqrt{2w}$ one obtains
$$
v_t=u_x, ~~~ uu_t=v_x+u_y,
$$
which, on elimination of $v$, leads to the dKP equation $(u_y-uu_t)_t+u_{xx}=0$, with  `non-standard'  notation for the independent variables.  The corresponding KP equation, $(u_y-uu_t)_t+u_{xx}+\epsilon^2u_{tttt}=0$, gives rise to the following integrable deformation of the original system:
  $$
  v_t=(\sqrt{2w})_x, ~~~ w_t=v_x+(\sqrt {2w})_y+\epsilon^2(\sqrt{2w})_{ttt}.
  $$
This, however,  is clearly outside the class of Hamiltonian deformations.

\section{Deformable Hamiltonians of type I}

In this section we summarise our results on deformations of Hamiltonian systems of the form
$$
\left(
  \begin{array}{c}
  v \\
    w \\
  \end{array}
\right)_t    =
\left(
  \begin{array}{cc}
    d/dx & 0 \\
    0 & d/dy \\
  \end{array}
\right) \left(
                                            \begin{array}{c}
                                              h_v \\
                                              h_w \\
                                            \end{array}
                                          \right),
$$
or,  explicitly,
$$
v_t=(h_v)_x, ~~~ w_t=(h_w)_y.
$$
The integrability  conditions constitute a system of fourth order PDEs for the Hamiltonian density $h(v, w)$ \cite{Fer1}:
\begin{equation}
\begin{array}{c}
h_{vw}(h_{vw}^2-h_{vv}h_{ww})h_{vvvv}=4h_{vw}h_{vvv}(h_{vw}h_{vvw}-h_{vv}h_{vww}) 

\\
+3h_{vv}h_{vw}h^2_{vvw}-2h_{vv}h_{ww}h_{vvv}h_{vvw}-h_{vw}h_{ww}h^2_{vvv}, \\
\ \\
h_{vw}(h_{vw}^2-h_{vv}h_{ww})h_{vvvw}=-h_{vw}h_{vvv}(h_{vv}h_{www}
+h_{ww}h_{vvw})\\
+3h^2_{vw}h^2_{vvw}-2h_{vv}h_{ww}h_{vvv}h_{vww}+h^2_{vw}h_{vvv}h_{vww}, \\
\ \\
h_{vw}(h_{vw}^2-h_{vv}h_{ww})h_{vvww}=4h^2_{vw}h_{vvw}h_{vww} \\
-h_{vv}h_{vvw}(h_{vw}h_{www}+h_{ww}h_{vww})-h_{ww}h_{vvv}(h_{vw}h_{vww}+h_{vv}h_{www}), 

\\
\ \\
h_{vw}(h_{vw}^2-h_{vv}h_{ww})h_{vwww}=-h_{vw}h_{www}(h_{ww}h_{vvv}+h_{vv}h_{vww}) 

\\
+3h^2_{vw}h^2_{vww}-2h_{vv}h_{ww}h_{www}h_{vvw}+h^2_{vw}h_{www}h_{vvw}, \\
\ \\
h_{vw}(h_{vw}^2-h_{vv}h_{ww})h_{wwww}=4h_{vw}h_{www}(h_{vw}h_{vww}-h_{ww}h_{vvw}) 

\\
+3h_{ww}h_{vw}h^2_{vww}-2h_{vv}h_{ww}h_{www}h_{vww}-h_{vw}h_{vv}h^2_{www}. \\
\end{array}
\label{44}
\end{equation}
This system is in involution, and is invariant under the $8$-parameter  group of Lie-point symmetries, 
\bea
&&{\nonumber}v \rightarrow av +  b,  \\
&&{\nonumber}w \rightarrow  cw + d,  \\
&&{\nonumber}h \rightarrow \alpha h + \beta v + \gamma w +\delta,
\eea
which constitute the equivalence group of the problem. Furthermore, there is an obvious symmetry corresponding to the interchange of $v$ and $w$. Particular solutions include
$$
h(v, w)=vw+\alpha v^3, ~~~ h(v, w)=w\sqrt v+\alpha v^{5/2}, ~~~ h(v, w)=\frac{1}{2}(v+w)^2+e^v, ~~~ h(v, w)=(w+a(v))^2,
$$
where $a(v)$ solves the ODE $a'a''a''''=2a''^2a'''+a'a'''^2$, etc, see \cite{FMS} for the general discussion. As before, generic Hamiltonians are not deformable. Although Case I  turns out to be  considerably more complicated from computational point of view, our calculations support the conjecture that deformable densities of type I are characterised by exactly the same additional  constraints as in case II:

\medskip

\noindent {\bf Conjecture} {\it  A Hamiltonian  $H_0=\int h(v, w)\ dxdy $  of type I  possesses a nontrivial integrable deformation to the order $\epsilon^2$ if and only if, along with the integrability conditions (\ref{44}),   it satisfies the additional differential constraints
$$
h_{vvv}h_{vww}-h_{vvw}^2=0, ~~ h_{vvv}h_{www}-h_{vvw}h_{vww}=0, ~~ h_{www}h_{vvw}-h_{vww}^2=0,
$$
or, equivalently, 
$$
{\rm rank} \ \left(
\begin{array}{ccc}
h_{vvv} & h_{vvw} & h_{vww} \\
h_{vvw} & h_{vww} & h_{www} 
\end{array}
\right)=1.
$$}

\medskip

\noindent Modulo equivalence transformations, this gives three types of deformable densities:
$$
h(v, w)=vw+\alpha v^3, ~~~~  h(v, w)=\frac{1}{2}(v+w)^2+e^v, ~~~ h(v, w)=\frac{\alpha}{2}v^2+\beta vw+\frac{\gamma}{2}w^2+f(v+w),
$$
where $f$ satisfies the  integrability condition 
$$
(\beta+f'')\triangle f''''= f'''^2[3\triangle+(\beta-\alpha)(\beta-\gamma)],
$$
here $\triangle=(\beta+f'')^2-(\alpha+f'')(\gamma+f'')$.
Dispersive deformations of these Hamiltonians are given by the following formulae (we use the normalisation $f_8=f_9=0$ which can always be achieved by a canonical transformation; furthermore, all $\epsilon$-deformations are trivial, and have been set equal to zero):
 $$
 H=\int  vw+\alpha v^3+{\epsilon^2}(6\alpha v_x^2+v_xw_x)+O(\epsilon^4)\ dxdy,
 $$
$$
 H=\int \frac{1}{2}(v+w)^2+e^v+{\epsilon^2}e^v\left(2(1+e^v) v_x^2-2w_x^2-2e^v v_xw_x+v_xw_y+v_yw_x\right)+O(\epsilon^4)\ dxdy.
 $$
The third case is somewhat more complicated: 
$$
 H=\int \frac{\alpha}{2}v^2+\beta vw+\frac{\gamma}{2}w^2+f(v+w)+{\epsilon^2}h_2+O(\epsilon^4)\ dxdy,
 $$
where $h_2=f_1v_x^2+f_2v_y^2+f_3w_x^2+f_4w_y^2+f_5v_xw_x+f_6(v_xw_y+v_yw_x)+f_7v_yw_y$, and the coefficients $f_1 - f_7$ are defined as follows:
$$
f_1=(\alpha+f'')(\beta+f'')^2\triangle, ~~~ f_4=(\gamma+f'')(\beta+f'')^2\triangle,
$$
$$
f_2=(4\beta-\alpha+3f'')(\beta+f'')^2\triangle, ~~~
f_3=(4\beta-\gamma+3f'')(\beta+f'')^2\triangle,
$$
$$
f_5=(\beta+f'')\triangle[\triangle+4(\alpha+f'')(\beta+f'')], ~~~ f_7=(\beta+f'')\triangle[\triangle+4(\gamma+f'')(\beta+f'')],
$$
$$
f_6=\frac{1}{2}(\beta+f'')\triangle[2\triangle+(2\beta-\alpha-\gamma)^2]. 
$$
We conjecture that  Hamiltonians from Theorem 3 can be deformed to all orders in $\epsilon$.

\section{Deformable Hamiltonians of type III}

In this section we consider Hamiltonian systems of the form
$$
\left(
  \begin{array}{c}
  v \\
    w \\
  \end{array}
\right)_t    =
\Bigg [ \left(
\begin{array}{cc}2v&w\\
w & 0
\end{array}
\right)\frac{d}{dx}+
 \left(
\begin{array}{cc}
0&v\\
v & 2w
\end{array}
\right)\frac{d}{dy}+
 \left(
\begin{array}{cc}
v_x&v_y\\
w_x & w_y
\end{array}
\right)\Bigg ]
 \left(
                                            \begin{array}{c}
                                              h_v \\
                                              h_w \\
                                            \end{array}
                                          \right),
$$
or,  explicitly,
$$
v_t=(2vh_v+wh_w-h)_x+(vh_w)_y, ~~~ w_t=(wh_v)_x+(2wh_w+vh_v-h)_y.
$$
The integrability conditions constitute a   system of fourth order PDEs for the Hamiltonian density $h(v, w)$  which is not presented here due to its complexity. We have verified in \cite{FOS} that this system is in involution, and its solution space is $10$-dimensional. It is invariant under an $8$-dimensional group  of Lie-point symmetries,
\bea
&&{\nonumber}v \rightarrow av +  bw,  \\
&&{\nonumber}w \rightarrow cv+ dw ,  \\
&&{\nonumber}h \rightarrow \alpha h + \beta v + \gamma w +\delta,
\eea
which constitute the equivalence group of the problem.
Particular integrable Hamiltonian densities include
$$
h(v, w)=\frac{w}{v}+\alpha v^2, ~~~ h(v, w)=\frac{w}{v}+(v+c)\ln (v+c), ~~~ h(v, w)=(vw)^{{1/3}}.
$$
A more complicated example has the form $h(v, w)=wf(v/w^2)$ where the function $f(y)$ solves the ODE
$$
y^2f'f''(f'+2yf'')f''''=2y^2f'(f'+3yf'')f'''^2
$$
$$
+2yf''(2f'^2+9yf'f''+4y^2f''^2)f'''+3f''^2(2f'^2+8yf'f''+5y^2f''^2).
$$
However,  calculations suggest that none of these examples are deformable:

\medskip

\noindent {\bf Conjecture} {\it For Hamiltonians of type III, all deformations of the order $\epsilon^2$ are trivial.}

\section{Concluding remarks}

In this paper we discuss, in the spirit of \cite{Dubrovin3} -- \cite{Dubrovin7},  the deformation theory of $2+1$ dimensional Hamiltonian systems of hydrodynamic type, defined by local Poisson brackets and local Hamiltonians.  Our results demonstrate that, already at the order $\epsilon^2$, the requirement of the existence of nontrivial dispersive deformations  is very restrictive so that `generic' integrable Hamiltonians are not deformable.  The main reason for this is apparently the assumption that all higher order dispersive corrections are  {\it local} expressions in the dependent variables $v, w$ and $x, y$-derivatives thereof.  It would be of interest to extend this scheme to the case of nonlocal brackets/Hamiltonians, see \cite{B} for particular examples obtained via Dirac reduction.  

Furthermore, to the best of our knowledge the theory of  deformations of  multi-dimensional Poisson brackets of hydrodynamic type has not been constructed: is it  true that all such deformations are trivial, as in the $1+1$ dimensional case?

Finally, calculations leading to Example 1 of Sect. 2 show that any Hamiltonian of the form  $H_0=\int \frac{w^2}{2}+f(v)\ dxdy$, where the function $f$ is arbitrary,  possesses  a unique dispersive  deformation of the form 
  $$
 H=\int  \frac{w^2}{2}+f(v)+\frac{\epsilon^2}{3!}f'''v_x^2+O(\epsilon^4)\ dxdy,
 $$
which inherits all one-phase solutions to the order $\epsilon^2$. Thus, one can speak of  `partial integrability' of a certain kind. However, already the requirement of the inheritance of two-phase solutions forces $f$ to satisfy the integrability condition  $f''''f''=f'''^2$.

\section*{Acknowledgements}

 We  thank Antonio Moro for his involvement at the early stage of this project when the role of $3$-component reductions in the proof of triviality of $\epsilon$-deformations was first understood. We also thank  Boris Dubrovin, John Gibbons  and Maxim Pavlov for clarifying discussions. The research of EVF  was partially supported by the European Research Council Advanced Grant  FroM-PDE.


\begin{thebibliography}{99}



\bibitem{Bla} M. Blaszak and B.M. Szablikowski, Classical R-matrix theory for bi-Hamiltonian field systems, J. Phys. A {\bf 42}, no. 40 (2009) 404002, 35 pp. 





\bibitem{B}  M. Blaszak and B.M.  Szablikowski,  Classical $R$-matrix theory of dispersionless systems. II.
$(2+1)$ dimension theory,  J. Phys. A {\bf 35},  no. 48 (2002) 10345-10364.



\bibitem{BF} C.P. Boyer and J.D. Finley, Killing vectors in self-dual Euclidean Einstein spaces, J. Math. Phys. {\bf 23} (1982) 1126-1130.



\bibitem{Carlet}  G.  Carlet, B. Dubrovin and L.P.  Mertens,  
Infinite-Dimensional Frobenius Manifolds for 2+1 Integrable Systems,   Math. Ann. {\bf 349}, no. 1 (2011) 75-115. 

\bibitem{Magri} L. Degiovanni, F.  Magri and V. Sciacca,  On deformation of Poisson manifolds of hydrodynamic type, Comm. Math. Phys. {\bf 253}, no. 1 (2005) 1-24. 


\bibitem{Dubrovin1} B.A. Dubrovin and S.P.  Novikov,  Poisson brackets of hydrodynamic type,  Dokl. Akad. Nauk
SSSR {\bf 279}, no. 2  (1984) 294-297.

\bibitem{Dubrovin2} B.A. Dubrovin and S.P.  Novikov, Hydrodynamics of weakly deformed soliton lattices. Differential geometry and Hamiltonian theory,   Russian Math. Surveys {\bf 44}, no. 6 (1989) 35-124.



\bibitem{Dubrovin3} B.A. Dubrovin, Hamiltonian PDEs: deformations, integrability, solutions, J. Phys. A {\bf 43}, no. 43 (2010) 434002, 20 pp.

\bibitem{Dubrovin4} B.A. Dubrovin,  Hamiltonian partial differential equations and Frobenius manifolds,  Russian Math. Surveys {\bf 63}, no. 6 (2008) 999-1010.


\bibitem{Dubrovin8} B.A. Dubrovin,  On universality of critical behaviour in Hamiltonian PDEs, in Geometry, topology, and mathematical physics, 
Amer. Math. Soc. Transl.  {\bf 224}, Ser. 2  (2008) 59-109. 

\bibitem{Dubrovin5} B.A. Dubrovin,  On Hamiltonian perturbations of hyperbolic systems of conservation laws. II. Universality of critical behaviour, Comm. Math. Phys. {\bf 267}, no. 1 (2006) 117-139.

\bibitem{Dubrovin6} B.A. Dubrovin,  Si-Qi Liu and Youjin Zhang,  On Hamiltonian perturbations of hyperbolic systems of conservation laws. I. Quasi-triviality of bi-Hamiltonian perturbations,  Comm. Pure Appl. Math. {\bf 59}, no. 4 (2006) 559-615.

\bibitem{Dubrovin7} B.A. Dubrovin and Youjin  Zhang, Bi-Hamiltonian hierarchies in 2D topological field theory at one-loop approximation,  Comm. Math. Phys. {\bf 198}, no. 2 (1998) 311-361.

\bibitem{Fer1} E.V. Ferapontov and K.R. Khusnutdinova, On integrability of (2+1)-dimensional quasilinear systems,
Comm. Math. Phys. {\bf 248} (2004) 187-206.


\bibitem{FMS} E.V. Ferapontov, A. Moro and  V. V. Sokolov, 
Hamiltonian systems of hydrodynamic type in 2+1 dimensions, Comm. Math. Phys. {\bf 285}, no. 1 (2009) 31-65.

\bibitem{FOS} E.V. Ferapontov, A.V. Odesskii and N.M. Stoilov, Classification of integrable two-component  Hamilonian systems of hydrodynamic type in 2+1 dimensions, Journal of Mathematical Physics, {\bf 52} (2011)  073505; Manuscript ID: 10-1152R.

\bibitem{FM} E.V. Ferapontov and A. Moro, 
Dispersive deformations of hydrodynamic reductions of 2D dispersionless integrable systems,  J. Phys. A: Math. Theor. {\bf 42}, no. 3 (2009) 035211, 15 pp.

\bibitem{FMN} E.V. Ferapontov, A. Moro and V.S. Novikov, Integrable equations in $2+1$ dimensions: deformations of dispersionless limits, J. Phys. A: Math. Theor. {\bf 42}, no. 34 (2009)  345205, 18 pp.



\bibitem{Getzler} E. Getzler, A Darboux theorem for Hamiltonian operators in the formal calculus of variations,  Duke Math. J. {\bf 111} (2002) 535-560.




\bibitem{GibTsa96} J. Gibbons and S.P. Tsarev, Reductions of the
Benney equations, Phys. Lett. A {\bf 211} (1996) 19-24.







\bibitem{Mokhov1} O.I.  Mokhov,  Poisson brackets of Dubrovin-Novikov type (DN-brackets), Funct. Anal. Appl.
{\bf 22}, no. 4 (1988) 336-338.

\bibitem{Mokhov2} O.I.  Mokhov,  Classification of non-singular multi-dimensional Dubrovin-Novikov brackets,  Funct. Anal. Appl. {\bf 42}, no. 1 (2008)  33-44.


\bibitem{Tsarev0} S.P. Tsarev,  Poisson brackets and one-dimensional Hamiltonian systems of hydrodynamic type, Dokl. Akad. Nauk SSSR {\bf 282}, no. 3 (1985) 534-537.

\bibitem{Tsarev} S.P. Tsarev, Geometry of Hamiltonian systems of
hydrodynamic type. The generalized hodograph method, Izvestija AN USSR
Math. {\bf 54}, no. 5  (1990) 1048-1068.


\end{thebibliography}
\end{document}